 \documentclass[final,5p,times,twocolumn]{elsarticle}

 \usepackage{graphics}
 \usepackage{graphicx}
\usepackage{amssymb}
\journal{Journal of Molecular Spectroscopy}

\begin{document}

\begin{frontmatter}

%% Title, authors and addresses

%% use the tnoteref command within \title for footnotes;
%% use the tnotetext command for the associated footnote;
%% use the fnref command within \author or \address for footnotes;
%% use the fntext command for the associated footnote;
%% use the corref command within \author for corresponding author footnotes;
%% use the cortext command for the associated footnote;
%% use the ead command for the email address,
%% and the form \ead[url] for the home page:
%%
%% \title{Title\tnoteref{label1}}
%% \tnotetext[label1]{}
%% \author{Name\corref{cor1}\fnref{label2}}
%% \ead{email address}
%% \ead[url]{home page}
%% \fntext[label2]{}
%% \cortext[cor1]{}
%% \address{Address\fnref{label3}}
%% \fntext[label3]{}

\title{The spectroscopic parameters of sodium cyanide, NaCN (\~X$^1$A'), revisited}

%% use optional labels to link authors explicitly to addresses:
%% \author[label1,label2]{<author name>}
%% \address[label1]{<address>}
%% \address[label2]{<address>}

\author[Koeln]{Holger S.P. M\"uller\corref{cor}}
\ead{hspm@ph1.uni-koeln.de}
\cortext[cor]{Corresponding author.}
\author[Tucson]{D.T. Halfen}
\author[Tucson]{L.M. Ziurys}

\address[Koeln]{I.~Physikalisches Institut, Universit{\"a}t zu K{\"o}ln, 
   Z{\"u}lpicher Str. 77, 50937 K{\"o}ln, Germany}
\address[Tucson]{Departments of Chemistry and Astronomy, Arizona Radio Observatory and 
  Steward Observatory, University of Arizona, 933~North Cherry Avenue, Tucson, AZ~85721, 
  USA}

%%%%%%%%%%%%%%%%%%%%%%%%%%%%%%%%%%%%%%%%%%%%%%%%%%%%%%%%%%%%%%%%%%%%%%%%%%%%%%%%%%%%%
%%%%%%%%%%%%%%%%%%%%%%%%%%%%%%%%%%%%%%%%%%%%%%%%%%%%%%%%%%%%%%%%%%%%%%%%%%%%%%%%%%%%%
%%%%%%%%%%%%%%%%%%%%%%%%%%%%%%%%%%%%%%%%%%%%%%%%%%%%%%%%%%%%%%%%%%%%%%%%%%%%%%%%%%%%%
\begin{abstract}

The study of the rotational spectrum of NaCN (\~X$^1$A') has recently been extended 
in frequency and in quantum numbers. Difficulties have been encountered in fitting 
the transition frequencies within experimental uncertainties. Various trial fits 
traced the difficulties to the incomplete diagonalization of the Hamiltonian. 
Employing fewer spectroscopic parameters than before, the transition frequencies 
could be reproduced within experimental uncertainties on average. 
Predictions of $a$-type $R$-branch transitions with $K_a \le 7$ up to 570~GHz should 
be reliable to better than 1~MHz. In addition, modified spectroscopic parameters 
have been derived for the $^{13}$C isotopic species of NaCN.

\end{abstract}

\begin{keyword}
%% keywords here, in the form: keyword \sep keyword

rotational spectroscopy \sep
circumstellar molecule \sep
centrifugal distortion \sep
Hamiltonian diagonalization

%% MSC codes here, in the form: \MSC code \sep code
%% or \MSC[2008] code \sep code (2000 is the default)

\end{keyword}

\end{frontmatter}

%%
%% Start line numbering here if you want
%%
% \linenumbers

%%%%%%%%%%%%%%%%%%%%%%%%%%%%%%%%%%%%%%%%%%%%%%%%%%%%%%%%%%%%%%%%%%%%%%%%%%%%%%%%%%%%%
%%%%%  main text  %%%%%%%%%%%%%%%%%%%%%%%%%%%%%%%%%%%%%%%%%%%%%%%%%%%%%%%%%%%%%%%%%%%
%%%%%%%%%%%%%%%%%%%%%%%%%%%%%%%%%%%%%%%%%%%%%%%%%%%%%%%%%%%%%%%%%%%%%%%%%%%%%%%%%%%%%

%%%%%%%%%%%%%%%%%%%%%%%%%%%%%%%%%%%%%%%%%%%%%%%%%%%%%%%%%%%%%%%%%%%%%%%%%%%%%%%%%%%%%
%%%%%  Introduction  %%%%%%%%%%%%%%%%%%%%%%%%%%%%%%%%%%%%%%%%%%%%%%%%%%%%%%%%%%%%%%%%
%%%%%%%%%%%%%%%%%%%%%%%%%%%%%%%%%%%%%%%%%%%%%%%%%%%%%%%%%%%%%%%%%%%%%%%%%%%%%%%%%%%%%

\section{Introduction}
\label{introduction}

There are several ways to form a molecule from a metal M and the CN group. 
Several metals, such as the transition metals copper and zinc, prefer a linear cyanide 
arrangement with an MC bond. Linear isocyanides with an MN bond are preferred by others 
such as the main group metals magnesium and aluminum. Sodium and potassium, however, 
display a different structural motif, a nearly T-shaped structure with almost equally 
long MC and MN bonds. As a consequence, both molecules are asymmetric top rotors.

The rotational spectrum of NaCN has been studied by van Vaals et al.~\cite{NaCN_rot_1984} 
between 9 and 40~GHz employing molecular-beam electric-resonance spectroscopy. 
20 rotational transitions were recorded, half of them being much weaker $b$-type 
transitions. NaCN was detected in the circumstellar envelope of the carbon-rich star 
CW~Leo, also known as IRC~+10216~\cite{det_NaCN_1994}, soon after three pairs of 
related unidentified lines, also detected in that source, had been attributed 
to MgNC~\cite{MgNC_rot_1993}. 
Recently, He et al.~\cite{IRC_10216_2008} described a molecular line survey of CW~Leo 
at 2 and 1.3~mm. In order to analyze the NaCN emission, the spectroscopic parameters 
of NaCN were reevaluated because initially, the quartic centrifugal distortion 
parameters were expressed as $\tau$ values, which are rarely implemented in current 
spectroscopy programs. The predictions derived from the resulting parameters were 
good enough for the NaCN emission features obtained in the course of that line survey 
as well as for previously published transitions. 
Observations at slightly higher frequencies and with a better signal-to-noise ratio, 
however, revealed increasing deviations largely with increasing $J$, prompting 
Halfen and Ziurys~\cite{NaCN_rot_2011} to study the rotational spectrum of NaCN 
between approximately 180 and 530~GHz. They encounterted difficulties in 
reproducing their experimental transition frequencies on average within experimental 
uncertainties and attributed this to the floppy nature of the molecule. 

In the course of creating an updated catalog entry for the Cologne Database for 
Molecular Spectroscopy (CDMS)~\cite{CDMS_1,CDMS_2}, the previously reported 
experimental data for NaCN~\cite{NaCN_rot_1984,NaCN_rot_2011} were critically 
evaluated. Using fewer spectroscopic parameters than before, it was possible to 
reproduce both data sets well if the Hamiltonian was diagonalized sufficiently.

%%%%%%%%%%%%%%%%%%%%%%%%%%%%%%%%%%%%%%%%%%%%%%%%%%%%%%%%%%%%%%%%%%%%%%%%%%%%%%%%%%%%%
%%%%%  Spectral analysis  %%%%%%%%%%%%%%%%%%%%%%%%%%%%%%%%%%%%%%%%%%%%%%%%%%%%%%%%%%%
%%%%%%%%%%%%%%%%%%%%%%%%%%%%%%%%%%%%%%%%%%%%%%%%%%%%%%%%%%%%%%%%%%%%%%%%%%%%%%%%%%%%%

\section{Considerations for the spectral analysis}
\label{considerations}

One finds rather frequently that the set of spectroscopic parameters, which reproduces 
a given list of transition frequencies within experimental uncertainties, is not unique. 
One way of reducing the ambiguity of a parameter set is to reduce the rms error, also 
knows as the weighted rms, of the fit with as few parameters as possible and reasonable. 
A reasonable parameter set requires certain lower order term to be used before one 
of higher order is allowed to be included; e.g., the inclusion of $L_{JJK}$ 
in the fit requires that at least one of the two parameters $H_J$ or $H_{JK}$ 
has already been used in the fit. In the absence of strong correlation among 
the parameters, it suffices to search for the one which reduces the rms error the most. 
However, such correlation is rather common, such that searches for small parameter sets 
may be more complex. It is frequently assumed that the smallest parameter set, 
which reproduces the parameter set best, will provide the most reasonable predictions.
It is useful in this context to address the issue of uncertainties attributed to the 
transition frequencies. Besides appropriate estimates and, less frequently in recent 
years, no estimates at all, one also finds uncertainties which seem to be too optimistic 
or too pessimistic, leading to weights in enlarged data sets which are too high or too low, 
respectively, if these estimates stay uncorrected. If the list of transition frequencies 
is sufficiently large with respect to the number of spectroscopic parameters varied 
in the fit, an rms error of 1.0 overall as well as for well defined sub-sets 
of the frequency list can be viewed as an indication of reasonable uncertainty estimates, 
though caution is advised if a very small set of the lines depends particularly strongly 
on one or few parameters. An additional criterion for reasonable uncertainty estimates 
is the distribution of the residuals. Most of the lines, about two third, should fit 
within the uncertainties, but there may be some lines with residuals exceeding three times 
the uncertainties, however, their number should decrease very rapidly with the magnitude 
of the ratio between residual and uncertainty. It is possible that the overall 
uncertainty estimates in a line list are too pessimistic, but one or few lines have 
residuals which are too large. A rejection threshold is applied in final fits 
in particular in larger data sets as a small set of transition frequencies with 
large residuals may complicate the determination of higher order parameters. 
A threshold of $3\sigma$ is rather common, see e.g. Ref.~\cite{SO2_rot_2005,DOBr_2010}, 
but other thresholds, usually less strict, may also be found.

%%%%%%%%%%%%%%%%%%%%%%%%%%%%%%%%%%%%%%%%%%%%%%%%%%%%%%%%%%%%%%%%%%%%%%%%%%%%%%%%%%%%%
%%%%%  Table 1?  %%%%%%%%%%%%%%%%%%%%%%%%%%%%%%%%%%%%%%%%%%%%%%%%%%%%%%%%%%%%%%%%%%%%
%%%%%%%%%%%%%%%%%%%%%%%%%%%%%%%%%%%%%%%%%%%%%%%%%%%%%%%%%%%%%%%%%%%%%%%%%%%%%%%%%%%%%

\begin{table*}
  \caption{Spectroscopic parameters$^a$ (MHz) of NaCN (\~X$^1$A') from the present combined fit in 
           comparison to fits employing microwave data only and in comparison to KCN along with 
           dimensionless rms errors of each fit.}
  \label{parameters}
{\footnotesize
  \begin{tabular}{lr@{}lr@{}lr@{}lr@{}lr@{}lr@{}l}
  \hline
   Parameter & \multicolumn{2}{l}{NaCN, combined fit} &  \multicolumn{2}{l}{Na$^{13}$CN} & 
\multicolumn{4}{l}{NaCN, microwave data only} & \multicolumn{2}{l}{KCN} \\
\cline{6-9}
    & \multicolumn{2}{l}{Present} & \multicolumn{2}{l}{Present} & \multicolumn{2}{l}{Ref.~\cite{IRC_10216_2008}} & 
\multicolumn{2}{l}{Present} & \multicolumn{2}{l}{Ref.~\cite{IRC_10216_2008}} \\
  \hline
$A - (B+C)/2$            & 50101&.74547~(110)   & 48021&.93283~(314)   & 50101&.74757~(166)    & 50101&.74606~(131)   &  53528&.08107~(268)   \\
$(B+C)/2$                &  7820&.189927~(145)  &  7652&.415049~(391)  &  7820&.190252~(259)   &  7820&.190351~(206)  &   4737&.751389~(93)   \\
$(B-C)/4$                &   274&.1524216~(124) &   272&.7985489~(396) &   274&.1525055~(177)  &   274&.1525120~(172) &    100&.959727~(64)   \\
$D_K \times 10^3$        &   194&.11~(25)       &   268&.89~(142)      &   196&.52~(169)       &   193&.56~(69)       &    956&.79~(247)      \\
$D_{JK} \times 10^3$     &   800&.06~(2)        &   738&.90~(39)       &   800&.23~(34)        &   800&.24~(19)       &    374&.609~(50)      \\
$D_J \times 10^3$        &    13&.3123~(5)      &    13&.2279~(180)    &    13&.3361~(160)     &    13&.3313~(96)     &      5&.24462~(209)   \\
$d_1 \times 10^3$        &  $-$2&.34537~(23)    &  $-$2&.42239~(112)   &  $-$2&.34908~(34)     &  $-$2&.34914~(33)    &   $-$0&.51452~(78)    \\
$d_2 \times 10^3$        &  $-$1&.407340~(56)   &  $-$1&.379543~(295)  &  $-$1&.408951~(182)   &  $-$1&.405461~(226)  &   $-$0&.237559~(199)  \\
$H_K \times 10^3$        &      &$-$            &      &$-$            &     1&.428~(220)      &       &$-$           &       &$-$            \\
$H_{KJ} \times 10^6$     &   327&.4~(18)        &   303&.5~(55)        &   155&.9~(184)        &    314&.5~(28)       &    101&.2~(11)        \\
$H_{JK} \times 10^6$     & $-$26&.31~(8)        & $-$23&.43~(30)       &  $-$3&.54~(120)       &  $-$26&.72~(19)      &   $-$7&.28~(11)       \\
$H_J \times 10^9$        &  $-$3&.77~(34)       &  $-$3&.5$^b$         &     0&.54~(5)         &       &$-$           &       &$-$            \\
$h_1 \times 10^9$        & $-$35&.68~(17)       & $-$34&.0$^b$         &      &$-$             &       &$-$           &       &$-$            \\
$h_2 \times 10^9$        & $-$25&.40~(9)        & $-$24&.6$^b$         &      &$-$             &  $-$47&.65~(199)     &   $-$5&.46~(74)       \\
$h_3 \times 10^9$        &    19&.34~(9)        &    19&.1$^b$         &    39&.92~(688)       &       &$-$           &       &$-$            \\
$L_{KKJ} \times 10^9$    &  $-$0&.384~(40)      &  $-$0&.331$^b$       &      &$-$             &       &$-$           &       &$-$            \\
$L_{JJK} \times 10^9$    &  $-$0&.845~(43)      &  $-$0&.759$^b$       &      &$-$             &       &$-$           &       &$-$            \\
$l_4 \times 10^{12}$     &  $-$0&.587~(80)      &  $-$0&.575$^b$       &      &$-$             &       &$-$           &       &$-$            \\
rms error                &     0&.746           &     0&.807           &     0&.840            &      1&.055          &      0&.827           \\
    \hline \hline
  \end{tabular}\\[2pt]
}
$^a$\footnotesize{Numbers in parentheses are one 
  standard deviation in units of the least significant figures.}\\
$^b$\footnotesize{Kept fixed to values from the combined NaCN fit multiplied 
  with ratios of the appropriate powers of $A - (B + C)/2$, $B + C$, and 
  $B - C$, see section~\ref{analysis}.}\\
\end{table*}

%%%%%%%%%%%%%%%%%%%%%%%%%%%%%%%%%%%%%%%%%%%%%%%%%%%%%%%%%%%%%%%%%%%%%%%%%%%%%%%%%%%%%
%%%%%%%%%%%%%%%%%%%%%%%%%%%%%%%%%%%%%%%%%%%%%%%%%%%%%%%%%%%%%%%%%%%%%%%%%%%%%%%%%%%%%
%%%%%%%%%%%%%%%%%%%%%%%%%%%%%%%%%%%%%%%%%%%%%%%%%%%%%%%%%%%%%%%%%%%%%%%%%%%%%%%%%%%%%

\section{Spectral analysis}
\label{analysis}

As mentioned in the introduction, NaCN as well as KCN are asymmetric tops, in contrast 
to most other molecules formed from a metal atom and the CN group. Since for both 
the N atom is closer to the metal atom~\cite{NaCN_rot_1984,KCN_rot_1984}, they should 
be viewed as isocyanides rather than cyanides. However, following the convention, 
we will continue calling these molecules cyanides. Ray's asymmetry parameter $\kappa$ 
is $-$0.9567 for NaCN, the molecule is thus fairly close to the prolate symmetric limit. 
It has a large $a$-dipole moment component of $\mu _a = 8.85$~D and a much smaller 
$b$-component of 
$\sim$0.2~D~\footnote{https://cdms.astro.uni-koeln.de/cgi-bin/cdmsinfo?file=e049510.cat}
\cite{IRC_10216_2008}. In spite of the small $b$-component, van Vaals et al. 
reported 10 $b$-type transitions with $\Delta J = \pm1$, $\Delta K_a = \mp1$, and 
$\Delta K_c = \pm1$ along with 10 $a$-type transitions with $\Delta J = 0, \pm1$, 
$\Delta K_a = 0$, and $\Delta K_c = \pm1$; the quantum numbers extended to $J$ and 
$K_a$ of 16 and 3, respectively. The $b$-type transitions contained essentially all 
of the $\Delta K_a \ne 0$ information in that data set as well as by far the most 
$\Delta J \ne 0$ information because most of the $a$-type transitions followed 
$\Delta J = 0$ selection rules. Inspection of their residuals suggested that the 
reported uncertainties are rather conservative; they were assumed to be 3$\sigma$ 
uncertainties in Ref.~\cite{IRC_10216_2008}. Employing thirteen parameters and the 
adjusted uncertainties, the latter authors were able to reproduce these lines 
with an rms error of 0.840. The fairly large number of parameters compared to 
the number of lines make it less clear whether the adjusted uncertainties of 
the transition frequencies were estimated appropriately. Nevertheless, the data 
permitted good to reasonable predictions well into the millimeter wave region.

Very recently, Halfen and Ziurys published an extensive set of NaCN transition 
frequencies all pertaining to the strong $a$-type $R$-branch transitions with 
$11 \le J'' \le 32$ and $0 \le K_a \le 6$ between 179 and 531~GHz~\cite{NaCN_rot_2011}. 
In order to provide an updated entry for the CDMS~\cite{CDMS_1,CDMS_2}, these data 
were subjected to combined fits together with the previous data~\cite{NaCN_rot_1984}. 
Since Halfen and Ziurys encountered difficulties in reproducing transitions mainly 
with higher $K_a$ values, it was decided to add the new transition frequencies 
with each $K_a$ series separately in the present analysis. Moreover, a tight rejection 
threshold of $3\sigma$ was applied since it seemed plausible that large residuals 
of one or few lines may have caused the difficulties. Pickett's {\scriptsize SPCAT} 
and {\scriptsize SPFIT} programs \cite{Herb} were used for prediction and fitting 
of the spectra, respectively, as has been done in the two previous 
investigations~\cite{IRC_10216_2008,NaCN_rot_2011}.

Starting with the parameter set of Ref.~\cite{IRC_10216_2008}, the previous transition 
frequencies plus 12 new ones with $K_a = 0$ could be reproduced overall within 
uncertainties by adding $h_2$ and omitting $H_K$ and $h_3$. At this stage, only one 
of the previous lines deviated from the calculated frequency by slightly more than 
three times the uncertainty. In later fits, the amount and identity of the lines 
displaying larger residual changed sometimes.

The sequential addition of transition frequencies having $K_a = 1 - 5$ afforded one 
new parameter each to be included. These were $h_1$, $h_3$, $L_{JJK}$, $l_4$, and 
$L_{KKJ}$. The addition of $K_a = 6$ lines did not require any new parameter.
Trial fits with $H_K$ included in the fit yielded a value of $41 \pm 55$~Hz with 
changes in the remaining parameters within the uncertainties. The uncertainties of 
the predominantly $K$-dependent terms increased markedly, but in no case by more 
than a factor of 2. Since $H_K$ was not determined with significance, 
it was omitted from the final fit. It is noteworthy that the uncertainty is a factor 
of 4 smaller than the uncertainty derived in Ref.~\cite{IRC_10216_2008} from the 
microwave data only.

This parameter set is smaller than the one in Ref.~\cite{NaCN_rot_2011}, but reproduces 
almost all of the data within the rejection threshold. Two of the new lines had residuals 
around 200~kHz, and one line deviated from its calculated position by more than 300~kHz. 
Most of the other lines seemed to have been judged conservatively with an uncertainty 
of 50~kHz because transition frequencies with $K_a \le 3$ gave partial rms errors of 
less than 0.5; the values for $K_a = 4$ and 6 were around 0.6, and only lines with 
$K_a = 5$ yielded a partial rms error of 1.19. 
The previous data~\cite{NaCN_rot_1984} now yielded a partial rms error of 1.17, 
indicating that the previously reported uncertainty estimates have been corrected 
slightly too far. But since none of these lines showed particularly large residuals 
with respect to the uncertainties, none of these lines were omitted and the 
uncertainties were left unchanged. 

The final set of spectroscopic parameters is given in Table~\ref{parameters} together 
with NaCN and KCN values from Ref.~\cite{IRC_10216_2008} and results from a fit 
of the NaCN microwave data to the KCN parameter set. The latter fit has 2 fewer 
parameters than the one published, but has a larger rms error of 1.055 compared 
with 0.840; this was the reason to discard this fit and to choose the one which 
required more parameters.

The list of transition frequencies with assignments, uncertainties, and residuals 
between experimental frequencies and those calculated from the final set of 
spectroscopic parameters is given in the Supplementary material. 
Predictions of the rotational spectrum along with a documentation will be provided 
in the catalog section\footnote{https://cdms.astro.uni-koeln.de/classic/entries/} 
of the CDMS. Line, parameter, and fit files, along with further auxiliary files, 
will be available in the archive 
section\footnote{https://cdms.astro.uni-koeln.de/classic/entries/archive/NaNC/}.

Ref.~\cite{IRC_10216_2008} also provided spectroscopic parameters for Na$^{13}$CN. 
Since predictions for the main isotopic species from that study turned out to be 
unsatifactory in the upper millimeter and lower submillimeter wave regions, it is 
rather likely that this applies to the predictions for Na$^{13}$CN also. 
17 rotational transitions had been studied by van Vaals et al.~\cite{NaCN_rot_1984}. 
The quoted uncertainties for the derived hyperfine free transition frequencies 
were assumed to be 3\,$\sigma$ values, as has been done for the main isotopolog. 
Centrifugal distortion parameters were estimated by multiplying the NaCN values from 
the combined fit with ratios of the appropriate powers of $A - (B + C)/2$, $B + C$, 
and $B - C$; e.\,g., $H_{KJ}$ with the ratio of $(A - (B + C)/2)^2(B + C)$ and $h_2$ with 
the ratio of $(B + C)(B - C)^2$. While this scaling can not be expected to hold strictly, 
it is usually a reasonable asssumption, see e.\,g. Ref.~\cite{IRC_10216_2008} for 
isotopic species of SiC$_2$ and for Na$^{13}$CN or Ref.~\cite{13C-VyCN-det_Sgr_B2} 
for the example of the isotopic species of vinyl cyanide. 

An rms error of 1.518 was achieved by just varying the rotational and quartic 
centrifugal distortion parameters. Significant reduction of the rms error 
below values of 1.0 were obtained when either $H_{KJ}$, $H_{JK}$, $H_J$, or $h_3$ 
were varied in addition, however, the changes for the latter two appeared to be 
too large in comparison to the respective values and were discarded. 
A slightly better fit with an rms error of 0.807 was obtained when both $H_{KJ}$ 
and $H_{JK}$ were released in the fit in addition to the lower order parameters. 
The resulting fit is also given in Table~\ref{parameters}. Updated predictions 
for Na$^{13}$CN as well as auxiliary files will be provided in the CDMS. 

%%%%%%%%%%%%%%%%%%%%%%%%%%%%%%%%%%%%%%%%%%%%%%%%%%%%%%%%%%%%%%%%%%%%%%%%%%%%%%%%%%%%%
%%%%%  Discussion  %%%%%%%%%%%%%%%%%%%%%%%%%%%%%%%%%%%%%%%%%%%%%%%%%%%%%%%%%%%%%%%%%%
%%%%%%%%%%%%%%%%%%%%%%%%%%%%%%%%%%%%%%%%%%%%%%%%%%%%%%%%%%%%%%%%%%%%%%%%%%%%%%%%%%%%%

\section{Discussion}
\label{Discussion}

It may be surprising that in the present study a better fit was obtained with 
17 parameters which is less than 23 employed previously~\cite{NaCN_rot_2011}. 
The reason turned out to be rather mundane. The programs {\scriptsize SPFIT} and 
{\scriptsize SPCAT} require minimum and maximum values for $K$ to be specified. 
In a prolate type, asymmetric top molecule, the minimum value is obviously 0, and 
the maximum value has to be at least as large as the highest $K_a$ value in the 
line list. The numerical diagonalization of the Hamiltonian, however, usually 
requires the maximum value to be somewhat larger than that value. 
It appears as if an increase by the highest $K_a$ order of the parameters 
in the fit is a reasonable estimate for this increase. 
Transitions in the NaCN line list extend to $K_a = 6$, and the parameter $l_4$ 
connects levels differing in $K_a$ by 8, e.g. the $K_a = 4$ asymmetry components. 
Hence, the maximum $K$ value requested in the fit should be around 14. In fact, no 
change in parameters and in the rms error occurs for $K_a \ge 15$. The data set for 
SO$_2$ in its ground vibrational state extends to $K_a = 28$~\cite{SO2_rot_2005}; 
trial fits have shown that a maximum $K$ value of at least 33 should be requested. 
Even larger maximum $K$ values may have to be requested for predicting a spectrum 
at room temperature for convergence of the calculated partition function. 

As can be seen in Table~\ref{parameters}, the lower order spectroscopic parameters, 
up to the quartic centrifugal distortion terms, agree very well with the previous 
parameter determined from microwave data only. Since merely selected parameters of 
higher order were employed, in particular for the MW fit, it is probably not surprising, 
that the agreement for the higher order parameters is not so good. In the alternative 
MW data fit, which employs exactly the parameters which were used in the KCN fit, 
the parameters $H_{KJ}$ and $H_{JK}$ agree well with the present values, and only 
$h_2$ differs by almost a factor of two.  It is interesting to note, that this fit 
yields much better predictions of the new data~\cite{NaCN_rot_2011} than the one 
published earlier~\cite{IRC_10216_2008}. The $19_{0,19} - 18_{0,18}$ transition, 
e.g., predicted by the alternative MW fit at 284160.864~(461)~MHz was measured 
at 284161.701~(50)~MHz, almost within twice the predicted uncertainty. 
The previously published fit \cite{IRC_10216_2008} predicts 284169.538~(634)~MHz. 
Apparently, the parameter set which yields the smallest rms error is 
not always the best one, and this situation is more likely to occur 
if the number of parameters is comparatively large with respect to the number of 
transition frequencies or if a certain parameter is determined by only few 
transitions. 

The present parameter set not only reproduces the experimental transitions well, 
but it should also permit to a certain degree reliable predictions beyond the 
experimentally accessed quantum numbers. It should be pointed out that the 
discussion of the predictive power of the present parameter set is concerned 
mainly with $a$-type transitions because the $a$-component of 8.85~D is very 
much larger than the $b$-component of $\sim$0.2~D. The $b$-type transitions 
are weaker than the $a$-type transitions by $(\mu _a/\mu _b)^2 \approx 1960$, 
making them unimportant for radioastronomy. No $b$-type transitions have been 
provided in the previous CDMS catalog entry, and this will also apply 
for the new entry.

We consider a prediction to be reliable if the frequency of a transition 
not yet measured will be found to deviate from the predicted frequency 
by an amount generally not exceeding three times the predicted uncertainty 
and never exceeding that value by far. Assuming that the effects of centrifugal 
distortion parameters not included in the fit do not exceed the predicted 
uncertainties, predictions may be reliable as long as the uncertainties 
do not exceed 1~MHz. At low values of $K_a$ this corresponds to $J$ up to 
around 50. Extrapolation in $K_a$ is very $J$-dependent. Around $J = 15$, 
$K_a = 10$ may be predicted reasonably, around $J = 35$, $K_a = 9$ is 
the limit, and beyond $J$ of 40, even predictions with $K_a = 8$ should be 
viewed with caution. Since the assumption mentioned above is not necessarily 
justified, all extrapolations should be viewed with additional caution.

The parameter $H_K$ may be the most important one not yet included in the fit 
which will have effects on the calculated transition frequencies and energies. 
Using a value of $H_K = 200$~Hz as fixed, a value almost four times the uncertainty 
of $H_K$ from trial fits, see section~\ref{analysis}, and thus near the upper end 
of possible values, changes the energies by less than 1~cm$^{-1}$ for $J = K_a \le 21$. 
The $a$-type transitions frequencies change by insignificant amounts for the 
$J$ and $K_a$ range for which the predictions are deemed to be reliable, see above. 

The easiest way to improve the $K$-level structure of NaCN will be recording 
higher-$J$ $a$-type $R$-branch transitions with $\Delta K_a = 0$ because of 
their favorable intensities. The $K_c = J$ transition are already approaching 
oblate pairing as the $J = 33 - 32$ transition frequencies differ by only 60~MHz. 
Transitions approaching oblate pairing contribute to the determination of 
purely $K$-dependent parameters. At still higher $J$, these transitions will 
reach oblate pairing and transitions with $K_c = J - 1$ etc. will approach and 
eventually reach oblate pairing also. In fact, the uncertainty of $H_K$ in 
present trials fits was reduced to 55~Hz (see section~\ref{analysis}), 
a factor of 4 smaller than the value of 220~Hz obtained from microwave data 
only \cite{IRC_10216_2008}, see also Table~\ref{parameters}, indicating 
that the newly measred $a$-type $R$-branch transitions \cite{NaCN_rot_2011} 
have improved the purely $K$-dependent parameters.

Obviously, $\Delta K_a \ne 0$ transitions probe the $K$-level structure more 
directly. In that case, it may be more promising to search for $\Delta K_a = 2$ 
transitions which are frequently much stronger than $b$-type transitions, 
in particular in the millimeter wave region. Transitions with $K_a \le 6$ 
and not too high $J$ should be predicted sufficiently well with the largest 
deviations from the predicted frequencies of order of 10~MHz.

The centrifugal distortion parameters which have been varied in the Na$^{13}$CN 
fit are close to the initial values obtained by scaling NaCN values. The most 
pronounced deviation occurs for $D_K$, which shows a very large isotopic shift. 
Such large shift, however, can be attributed to the accidentially small value 
of $D_K$ compared with $D_{JK}$ and is not uncommon. A large isotopic shift 
was also observed, e.\,g., for $\Delta _K$ of $^{37}$Cl$^{35}$ClO$_2$ compared to 
$^{35}$Cl$^{35}$ClO$_2$ which was fully accounted for by the harmonic force field 
calculation \cite{ClClO2_rot_1999}. A relatively much smaller deviation occurs 
for $d_1$, whose magnitude for the $^{13}$C isotopoic species is actually slightly 
larger than that of the main species. It is possible that this unexpected relation 
is caused by the assumptions made for some of the fixed higher order parameters.

Predictions of the rotational transition frequencies of Na$^{13}$CN should be viewed 
with much greater caution than those for the main isotopic species because of the 
much smaller data set. Nevertheless, the predictions should be reliable to 
reasonable at least up to $J$ of 15, maybe 20, and up to $K_a$ of 4, maybe 5.

Halfen and Ziurys~\cite{NaCN_rot_2011} derived a rotational temperature of about 
41~K for NaCN in the circumstellar envelope of CW~Leo. The Boltzmann peak at this 
temperature occurs slightly above 200~GHz, and the line intensities appear to be 
negligible above 500~GHz. Moreover, transitions with $K_a = 10$ are around 3 orders 
of magnitude weaker than those with $K_a = 0$ or 1. The predictions of the NaCN 
rotational spectrum derived from the present study should thus be adequate for all 
observational purposes, at least as far as the ground vibrational state is concerned. 
The bending mode ($\varv_2 = 1$) is the excited vibrational state lowest in energy. 
It is $\sim$170~cm$^{-1}$ or $\sim$245~K above ground~\cite{NaCN_rot_1984}. 
Transitions pertaining to this state will be hard to detect in the absence of 
infrared pumping, which, however, cannot be ruled out.

%%%%%%%%%%%%%%%%%%%%%%%%%%%%%%%%%%%%%%%%%%%%%%%%%%%%%%%%%%%%%%%%%%%%%%%%%%%%%%%%%%%%%
%%%%%  Conclusions  %%%%%%%%%%%%%%%%%%%%%%%%%%%%%%%%%%%%%%%%%%%%%%%%%%%%%%%%%%%%%%%%%
%%%%%%%%%%%%%%%%%%%%%%%%%%%%%%%%%%%%%%%%%%%%%%%%%%%%%%%%%%%%%%%%%%%%%%%%%%%%%%%%%%%%%

\section{Conclusions}
\label{Conclusions}

The rotational spectrum of NaCN has been reanalyzed. The previous difficulties in 
reproducing the experimental transition frequencies~\cite{NaCN_rot_2011} have been 
attributed to the incomplete diagonalization of the Hamiltonian. The predictions 
generated from the present data set should be good enough for all observational 
purposes restricted to the ground vibrational state. The modified parameter set 
for Na$^{13}$CN should provide better prediction of the rotational spectrum up 
to moderate quantum numbers.

%% The Appendices part is started with the command \appendix;
%% appendix sections are then done as normal sections
%% \appendix

%%%%%%%%%%%%%%%%%%%%%%%%%%%%%%%%%%%%%%%%%%%%%%%%%%%%%%%%%%%%%%%%%%%%%%%%%%%%%%%%%%%%%
%%%%%  acknowledgements  %%%%%%%%%%%%%%%%%%%%%%%%%%%%%%%%%%%%%%%%%%%%%%%%%%%%%%%%%%%%
%%%%%%%%%%%%%%%%%%%%%%%%%%%%%%%%%%%%%%%%%%%%%%%%%%%%%%%%%%%%%%%%%%%%%%%%%%%%%%%%%%%%%

\section*{Acknowledgements}

H.S.P.M. is very grateful to the Bundesministerium f\"ur Bildung und 
Forschung (BMBF) for financial support aimed at maintaining the 
Cologne Database for Molecular Spectroscopy, CDMS. This support has been 
administered by the Deutsches Zentrum f\"ur Luft- und Raumfahrt (DLR). 
L.M.Z. acknowledges support from NSF grant AST-09-06534.

\appendix

\section*{Appendix A. Supplementary material}

Supplementary data for this article are available on ScienceDirect (www.sciencedirect.com) 
and as part of the Ohio State University Molecular Spectroscopy Archives 
(http://library.osu.edu/sites/msa/jmsa\_hp.htm). Supplementary data associated with this 
article can be found, in the online version, at doi: 10.1016/j.jms.2011.12.005.

%% References
%%
%% Following citation commands can be used in the body text:
%% Usage of \cite is as follows:
%%   \cite{key}         ==>>  [#]
%%   \cite[chap. 2]{key} ==>> [#, chap. 2]
%%

%% References with bibTeX database:

\bibliographystyle{elsarticle-num}
\bibliography{<your-bib-database>}

%% Authors are advised to submit their bibtex database files. They are
%% requested to list a bibtex style file in the manuscript if they do
%% not want to use elsarticle-num.bst.

%% References without bibTeX database:

%%%%%%%%%%%%%%%%%%%%%%%%%%%%%%%%%%%%%%%%%%%%%%%%%%%%%%%%%%%%%%%%%%%%%%%%%%%%%%%%%%%%%
%%%%%%%%%%%%%%%%%%%%%%%%%%%%%%%%%%%%%%%%%%%%%%%%%%%%%%%%%%%%%%%%%%%%%%%%%%%%%%%%%%%%%
%%%%%%%%%%%%%%%%%%%%%%%%%%%%%%%%%%%%%%%%%%%%%%%%%%%%%%%%%%%%%%%%%%%%%%%%%%%%%%%%%%%%%

\end{document}